\documentclass[prl,twocolumn,showpacs,superscriptaddress,preprintnumbers,amssymb]{revtex4}
\usepackage{graphicx}
\usepackage{dcolumn}
\usepackage{bm}

\newcommand{\beq}{\begin{equation}}
\newcommand{\eeq}{\end{equation}}
\newcommand{\beqn}{\begin{eqnarray}}
\newcommand{\eeqn}{\end{eqnarray}}

\begin{document}

\title{Three dimensional Symmetry Protected Topological Phase close
to \\ Antiferromagnetic N\'{e}el order}


\author{Cenke Xu}

\affiliation{Department of Physics, University of California,
Santa Barbara, CA 93106}

\begin{abstract}

It is well-known that the Haldane phase of one-dimensional spin-1
chain is a symmetry protected topological (SPT) phase, which is
described by a nonlinear Sigma model (NLSM) with a $\Theta-$term
at $\Theta = 2\pi$. 
In this work we study a three dimensional SPT phase of SU(2$N$)
antiferromagnetic spin system with a self-conjugate representation
on every site. The spin ordered N\'{e}el phase of this system has
a ground state manifold $\mathcal{M} =
\frac{\mathrm{U}(2N)}{\mathrm{U}(N) \times \mathrm{U}(N)}$, and
this system is described by a NLSM defined with manifold
$\mathcal{M}$. Since the homotopy group $\pi_4[\mathcal{M}] =
\mathbb{Z}$ for $N > 1$, this NLSM can naturally have a
$\Theta-$term. We will argue that when $\Theta = 2\pi$ this NLSM
describes a SPT phase. This SPT phase is protected by the SU(2$N$)
spin symmetry, or its subgroup SU($N$)$\times$SU$(N)\rtimes Z_2$,
without assuming any other discrete symmetry. We will also
construct a trial SU(2$N$) spin state on a 3d lattice, we argue
that the long wavelength physics of this state is precisely
described by the aforementioned NLSM with $\Theta = 2\pi$.

\end{abstract}

\date{\today}

\maketitle



\section{ 1. Introduction}

According to the classic Ginzburg-Landau paradigm, all the
disordered phases of classical systems are basically equivalent
and completely featureless. However, it is now a consensus that
quantum disordered phases driven by quantum fluctuation can have
much richer structures. Roughly speaking, in quantum many-body
systems, quantum mechanics can lead to at least {\it three} types
of exotic/nontrivial quantum disordered phases: (1) Topological
phases with a gapped spectrum and bulk topological degeneracy, (2)
algebraic liquid phases with gapless bulk spectrum and power-law
correlations, and (3) symmetry protected topological phases. A
symmetry protected topological (SPT) phase is a state of matter
with gapped and nondegenerate bulk spectrum, but cannot
continuously evolve into a direct product state without a bulk
phase transition, when and only when the Hamiltonian of the entire
evolution is invariant under certain global symmetry
$G$~\cite{wenspt}. In terms of its phenomena, a SPT phase on a
$d-$dimensional lattice should satisfy at least the following
three criteria:

($1$). On a $d-$dimensional lattice without boundary, this phase
is fully gapped, and nondegenerate;

($2$). On a $d-$dimensional lattice with a $(d-1)-$dimensional
boundary, if the Hamiltonian of the entire system (including both
bulk and boundary Hamiltonian) preserves certain symmetry $G$,
this phase is either gapless, or gapped but degenerate.


(3). The boundary state of this $d-$dim system cannot be realized
as a $(d-1)$-dim lattice system built with the same onsite Hilbert
space, and with the same symmetry $G$.

If a $d-$dim quantum disordered phase satisfies all three criteria
(1), (2) and (3), this phase is a SPT phase. Both the $2d$ quantum
spin Hall (QSH) insulator~\cite{kane2005a,kane2005b,bernevig2006}
and $3d$ Topological band insulator
(TBI)~\cite{fukane,moorebalents2007,roy2007} are perfect examples
of SPT phases protected by time-reversal symmetry and charge
conservation.

Notice that the second criterion ($2$) implies the following two
possibilities: On a lattice with a boundary, the system is either
($2a$) gapless, or ($2b$) gapped but degenerate. When $d \geq 3$,
the degeneracy of ($2b$) can correspond to either spontaneous
breaking of $G$, or correspond to certain topological degeneracy
at the boundary. Which case occurs in the system will depend on
the detailed Hamiltonian at the boundary of the system. For
example, with interaction, the edge states of 2$d$ QSH insulator,
and 3$d$ TBI can both be gapped out through spontaneous
time-reversal symmetry breaking at the boundary, and this
spontaneous time-reversal symmetry breaking can occur through a
boundary transition, without destroying the bulk
state~\cite{xuedge,wuedge,xu3dedge}.

In this work we will focus on $bosonic$ spin systems. The simplest
example of SPT phase of spin system is the Haldane phase of one
dimensional spin-1 chain. In our paper we will first give a review
of Haldane phase, focusing on its nonlinear sigma model field
theory description in section II. Unlike the free fermion case,
although there is a classification of bosonic SPT using group
cohomology~\cite{wenspt}, specific models of higher dimensional
bosonic spin systems are not well understood. So far, in most
studies, construction of 2d and 3d bosonic SPTs has been focused
on systems with U(1)
symmetry~\cite{luashvin,levinsenthil,vishwanathsenthil}. In this
work we will study a $3+1$ dimensional analogue of the Haldane
phase, which is constructed as a SU(2$N$) spin state with a
self-conjugate representation on each site. Just like the Haldane
phase, this 3+1d SPT is described by a nonlinear sigma model
defined with a semiclassical antiferromagnetic order parameter
plus a topological $\Theta-$term.


\section{ 2. Haldane phase}

Although symmetry protected topological phase is a pure quantum
phenomenon without any classical analogue, the Haldane phase of
spin-1 chain 
can still be described semiclassically by a nonlinear Sigma model
(NLSM), which is defined $only$ in terms of the semiclassical
N\'{e}el order parameter $\vec{n}$~\cite{haldane1,haldane2}: \beqn
\mathcal{L} = \frac{1}{g} (\partial_\mu \vec{n})^2 +
\frac{i\Theta}{8\pi}\epsilon_{abc}\epsilon_{\mu\nu} n^a
\partial_\mu n^b \partial_\nu n^c. \label{o3theta} \eeqn When the
system has SO(3) symmetry, the entire manifold of the
configurations of N\'{e}el order parameter is $S^2$. If we assume
the trivial vacuum has $\Theta = 0$, then when $\Theta = 2\pi$,
this model describes the Haldane phase.

Haldane phase is a 1d SPT phase protected by SO(3) spin rotation
symmetry, namely as long as the SO(3) symmetry is preserved, no
other symmetry (such as time-reversal symmetry, reflection, etc.)
is required to protect the Haldane phase~\footnote{Actually the
SO(3) symmetry can be further relaxed, as discussed in
Ref.~\onlinecite{bergspt}, but for our purposes we will just
assume the SO(3) symmetry.}. This conclusion was established
through previous numerical simulations~\cite{bergspt}. The field
theory Eq.~\ref{o3theta} gives us the same conclusion, if it is
handled correctly.

The physical meaning of the $\Theta-$term in a NLSM is usually
interpreted as a factor $\exp(i\Theta)$ attached to every
instanton event in the space-time. Then this interpretation would
lead to the conclusion that $\Theta = 2\pi$ is equivalent to
$\Theta = 0$. However, this interpretation is very much
incomplete, because it only tells us that theories with $\Theta =
2\pi$ and $0$ have the same partition function when the system is
defined on a compact manifold. However, once we take an open
boundary condition in either space or time, the difference between
$\Theta = 2\pi$ and $0$ will be explicitly exposed. For example,
at the spatial boundary of the $1d$ system, $i.e.$ the interface
between $\Theta = 0$ and $2\pi$, the $\Theta-$term reduces to a
0+1$d$ O(3) Wess-Zumino-Witten (WZW) term at level 1, whose ground
state has two fold degeneracy, thus the boundary is effectively a
free spin-1/2 degree of freedom~\cite{ng1994}, which is exactly
the physics of Haldane phase. If we keep an open boundary at
temporal direction, then one can explicitly derive the ground
state wave function of Eq.~\ref{o3theta} at strong coupling, and
we can also see that the ground state of $\Theta = 2\pi$ and
$\Theta = 0$ are very different~\cite{xusenthil}.

We can also define time-reversal transformation: $Z_2^T : t
\rightarrow - t$, $\vec{n} \rightarrow - \vec{n}$, $i \rightarrow
-i$, Eq.~\ref{o3theta} is always invariant under $Z_2^T$ (notice
that $\tau = i t$ is invariant under $Z_2^T$), no matter which
value $\Theta$ takes. In fact, using the renormalization group
calculation(~\cite{pruisken1,pruisken2,pruisken3,pruisken4}, for a
more recent review, see Ref.~\onlinecite{pruisken5}), and the
general nonperturbative argument in Ref.~\onlinecite{xuludwig}, we
can derive a phase diagram for model Eq.~\ref{o3theta}: The system
is topological when $\Theta \in \left( (4k + 1)\pi, (4k+3)\pi
\right)$, while the system is trivial when $\Theta \in \left(
(4k-1)\pi, (4k+1)\pi \right) $; $\Theta = (2k+1)\pi$ is the
transition, where the bulk of the system is either gapless, or two
fold degenerate. Thus $\Theta = 0$ and $2\pi$ are two different
stable fixed points.

This phase diagram can be understood as follows:
The bulk partition function of Eq.~\ref{o3theta} is obviously
symmetric around $\Theta = 2\pi$ ($\Theta = 2\pi \pm \epsilon$
have the same partition function), thus $\Theta = 2\pi$ is a fixed
point that does not flow under RG. Tuning $\Theta $ slightly away
from $2\pi$ will not close the bulk gap, so it can only affect the
edge state. However, given that the boundary is a dangling
spin-1/2, then no perturbation can be added to the Hamiltonian
that can lift the spin-1/2 degeneracy at the boundary, as long as
the system has SO(3) symmetry, regardless of other discrete
symmetries. Thus if $\Theta$ is tuned slightly away from $2\pi$,
namely $\Theta = 2\pi \pm \epsilon$, as long as the system still
has SO(3) symmetry, the edge spin-1/2 doublet is still
stable~\cite{ng1994,xuludwig}. Thus $\Theta = 2\pi \pm \epsilon$
is in the same phase as $\Theta = 2\pi$. A similar effect was also
discussed in the context of 1+1d QED~\cite{coleman}. The edge
state can only be destroyed through a bulk transition, which
occurs at the transition $\Theta = \pi$. In this sense $\Theta =
2\pi$ is a stable fixed point of an entire Haldane phase. Thus the
Haldane phase is a SPT phase that requires SO(3) spin rotation
symmetry only.

Now let us couple two Haldane phases to each other: \beqn
\mathcal{L} &=& \frac{1}{g} (\partial_\mu \vec{n}_1)^2 + \frac{i
2\pi}{8\pi}\epsilon_{abc}\epsilon_{\mu\nu} n^a_1
\partial_\mu n^b_1 \partial_\nu n^c_1 \cr\cr &+& 1 \rightarrow 2 +
A (\vec{n}_1 \cdot \vec{n}_2). \label{2o3theta} \eeqn When $A = -
\infty$, effectively $\vec{n}_1 = \vec{n}_2 = \vec{n}$, then the
system is effectively described by one O(3) NLSM with $\Theta =
4\pi$; while when $A = + \infty$, the effective NLSM for the
system has $\Theta = 0$. When parameter $A$ is tuned from
$-\infty$ to $+\infty$, the entire phase diagram with $A \in
(-\infty, +\infty)$ is gapped in the bulk. Thus the theory with
$\Theta = 4\pi$ and $\Theta = 0$ are equivalent. This analysis
implies that with SO(3) symmetry, $1d$ spin systems have two
different classes: there is a trivial class with $\Theta = 4\pi
k$, and a nontrivial Haldane class with $\Theta = (4k + 2)\pi$.
This $Z_2$ classification is consistent with the group cohomology
formalism developed in Ref.~\cite{wenspt}.


There are various ways of describing the Haldane phase on a
lattice. In the follows we will choose one particular description
that will be generalized to higher dimensions later. The Haldane
phase can be described on a lattice as follows: On every site we
introduce a slave fermion with both spin-1/2 index and SU(2) color
index: $f_{i, A, \alpha}$, and the spin-1 operator is represented
as~\cite{xuspin1} \beqn \vec{S}_j = \frac{1}{2}\sum_{A = 1}^2
f^\dagger_{j, A,\alpha} \vec{\sigma}_{\alpha\beta} f_{j, A,\beta}.
\eeqn In order to match the slave fermion Hilbert space with the
spin-1 Hilbert space, we have to impose two different constraints
on each site: \beqn \sum_{\alpha, A} f^\dagger_{i, A, \alpha}
f_{i, A, \alpha} = 2,
 \ \ \ \sum_{\alpha, A, B} f^\dagger_{j, A, \alpha} \rho^\mu_{AB} f_{j,
B, \alpha}  = 0, \label{1dconstraint} \eeqn where $\rho^\mu$ are
three Pauli matrices of the color space. The second constraint
guarantees that on every site the color space is in a total
antisymmetric representation, thus the spin is in a total
symmetric spin-1 representation.

\begin{figure}
\begin{center} \includegraphics[width=3.2 in]{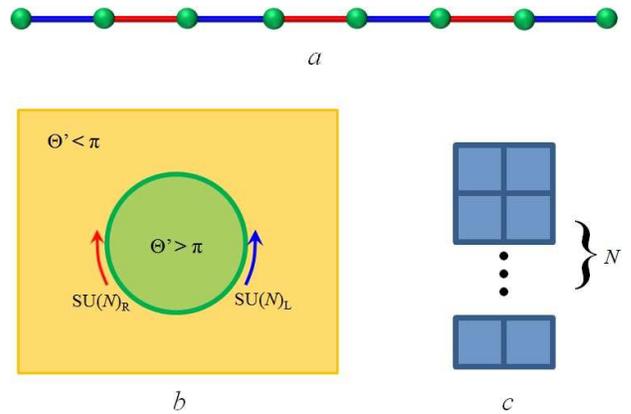}
\end{center}
\caption{ ($a$). The pictorial representation of the Haldane
phase, where the blue and red links stand for the valence bonds of
slave fermions with color $A = 1$ and $2$ respectively. ($b$) A
domain wall of $\Theta^\prime$ on the boundary of our 3$d$ SPT.
($c$) The Young diagram of the self-conjugate representation of
the SU(2$N$) spin system that we are considering.} \label{fig}
\end{figure}

The Haldane phase corresponds to the following mean field state of
$f_{A,\alpha}$: $f_{1,\alpha}$ forms valence bonds on links $(2i,
2i+1)$, while $f_{2, \alpha}$ forms valence bonds on links $(2i+1,
2i+2)$ (Fig.~\ref{fig}$a$). In terms of low energy field theory of
the slave fermion, the Haldane phase is described by the following
Lagrangian: \beqn \mathcal{L} = \bar{\psi} \gamma_\mu \partial_\mu
\psi + m_0 \bar{\psi} \rho^z \psi. \label{1dhaldanefermion} \eeqn
Here the Dirac fermion $\psi$ is the low energy mode of $f$, which
is expanded at the two Fermi points $k_f = \pm \pi/2$ in the 1d
Brillouin zone. If we couple the N\'{e}el order parameter to the
slave fermion, \beqn (-1)^j \vec{n}_j \cdot \sum_A f^\dagger_{j,
A, \alpha} \vec{\sigma}_{\alpha \beta} f_{j, A,\beta} \sim \vec{n}
\cdot \bar{\psi} \gamma_5 \vec{\sigma} \psi, \eeqn
Eq.~\ref{o3theta} can be derived after integrating out the slave
fermions~\cite{abanov2000}, and the derived $\Theta$ is precisely
$2\pi$. Notice that in Eq.~\ref{1dhaldanefermion} the gauge fields
introduced by constraints Eq.~\ref{1dconstraint} are ignored, but
in 1+1 dimension gauge fields are always confining, once the
matter fields are gapped.

The Haldane phase is a SPT phase {\it only} when the color-singlet
constraint $f^\dagger_j \rho^\mu f_j = 0$ is strictly imposed on
every site, $i.e.$ when the Hilbert space on every site is
rigorously spin-1. If this constraint is given up, the Hilbert
space on every site is enlarged to 6 dimension, and the Haldane
phase becomes trivial, because it can now be adiabatically
connected to a direct product state with spin-singlet on every
site. Actually, besides the Haldane phase mass gap $m_0$, we can
consider another mass gap $m_1 \bar{\psi} \gamma_5 \rho^z \psi$ of
the Dirac fermion $\psi$. Physically $m_1$ corresponds to a
``color density wave" on the lattice, which is not allowed if the
color singlet constraint is imposed strictly on every site.
Without the color singlet constraint, the Haldane phase can
adiabatically evolve into the color density wave state, by turning
on $m_1$.

In this section we have reviewed the physics of Haldane phase. For
Haldane phase we have both field theory description, and lattice
spin wave function. Most importantly, the field theory with
topological term can be precisely derived from the lattice wave
function. In the next section, we will achieve the same level of
understanding for a 3d generalization of Haldane phase.


\section{3. 3d SPT phase of SU(2$N$) spin system}

\subsection{3.1 Field Theory Description}

Let us try to look for higher dimensional generalizations of the
Haldane phase of spin-1 chain, which has a description in terms of
NLSM plus a $\Theta-$term. The most naive generalization would be
the AKLT state in higher dimensions, for instance the spin-2 AKLT
phase on the square lattice. The boundary of the spin-2 AKLT phase
on the square lattice is a spin-1/2 chain, which according to the
LSM theorem cannot be gapped and nondegenrate, thus the spin-2
AKLT state seems to be a SPT phase. However, in order to protect
the spin-2 AKLT state, we need translation symmetry, since the
boundary spin-1/2 chain can be dimerized and gapped out once the
translation symmetry of the system is explicitly broken. Thus this
is not an ideal generalization of the $1d$ Haldane phase, whose
stability does not rely on any translation symmetry. The spin-2
AKLT state on the square lattice can indeed be described by a NLSM
with a topological term, but the configurational space of this
NLSM would involve both the N\'{e}el and dimerization order
parameters.

The goal of this paper is to find a three dimensional SPT phase
{\it without} assuming the translation symmetry. Inspired by
Eq.~\ref{o3theta}, we should first look for magnetic systems,
whose ground state manifold $\mathcal{M}$ of the spin ordered
phase has a nontrivial homotopy group $\pi_4[\mathcal{M}] =
\mathbb{Z}$, then this SPT can be described by a NLSM defined in
manifold $\mathcal{M}$ with a $\Theta-$term. The SU(2$N$)
antiferromagnet with self-conjugate representation satisfies this
criterion: its magnetic ordered phase has
GSM~\cite{sachdev1990,sachdev1989nucl} \beqn \mathcal{M} =
\frac{\mathrm{U}(2N)}{\mathrm{U}(N) \times \mathrm{U}(N)}, \ \ \
\pi_4[\mathcal{M}] = \mathbb{Z}, \ \mathrm{for} \ N \geq 2. \eeqn
Every N\'{e}el order configuration $\mathcal{P} \in \mathcal{M}$
can be represented as \beqn \mathcal{P} = V^\dagger \Omega V, \ \
\ \Omega = \left(
\begin{array}{cccc}
\mathbf{1}_{N \times N}, & \mathbf{0}_{N \times N} \\ \\
\mathbf{0}_{N \times N}, & \mathbf{-1}_{N \times N}
\end{array}
\right) \eeqn where $V$ is a SU(2$N$) matrix. $\mathcal{P}$ is a
hermitian traceless order parameter that satisfies $\mathcal{P}^2
= \mathbf{1}$.
In fact, when $N = 1$, $\mathcal{M}$ is precisely $S^2$, and
$\mathcal{P}$ can always be represented as $\mathcal{P} =
\vec{n}\cdot \vec{\sigma}$, where $\vec{n}$ is the N\'{e}el
vector.

Since $\pi_4[\mathcal{M}] = \mathbb{Z}$, the following NLSM
defined on $\mathcal{M}$ can be written down: \beqn \mathcal{S}
&=& \int d^3x d\tau \ \frac{1}{g} \mathrm{tr}[\partial_\mu
\mathcal{P}
\partial_\mu \mathcal{P}] \cr\cr &+& \frac{i\Theta}{256\pi^2}
\mathrm{tr}[\mathcal{P} \partial_\mu \mathcal{P} \partial_\nu
\mathcal{P} \partial_\rho \mathcal{P} \partial_\lambda
\mathcal{P}] \epsilon_{\mu\nu\rho\lambda}. \label{3dp}\eeqn By
tuning the parameter $g$, there is obviously an order-disorder
transition. When $g $ is small, the system is in a spin ordered
phase where $\mathcal{P}$ is condensed and spontaneously breaks
the SU(2$N$) symmetry; when $g$ is large, the system is in a
disordered phase, and this disordered phase is what we are
interested in.

In the follows we will focus on the disordered phase of
Eq.~\ref{3dp} with $\Theta = 2\pi$, while assuming the trivial
vacuum of this spin system has $\Theta = 0$. Under the SU(2$N$)
transformation, order parameter $\mathcal{P}$ transforms as
$\mathcal{P} \rightarrow V^\dagger \mathcal{P} V$, where $V$ is a
SU(2$N$) matrix. Under time-reversal transformation, we take
$\mathcal{P}$ transform in the same way as the ordinary SU(2)
N\'{e}el order parameter: $\mathcal{P} \rightarrow -
\mathcal{P}^\ast$. Under this transformation, Eq.~\ref{3dp} and
Eq.~\ref{o3theta} are both invariant under time-reversal
transformation, no matter which value $\Theta$ takes. Thus
time-reversal symmetry is not required to protect $\Theta = 2\pi$.
We have argued that the Haldane phase does not need any discrete
symmetry (including time-reversal symmetry), as long as the SO(3)
symmetry is preserved. The same situation is true for the 3d SPT
phase discussed in this section: the stability of the 3d SPT phase
does not need time-reversal symmetry, as long as the SU(2$N$)
symmetry is preserved.

We will argue the quantum disordered phase of Eq.~\ref{3dp} is a
3d SPT phase when $\Theta = 2\pi$. Our argument proceeds in two
steps: $(1)$, the boundary of the system must be either gapless or
degenerate; $(2)$, the boundary cannot be realized as a 2d system
with the same symmetry as the bulk.

{\it Step 1: argue the edge state must be either gapless or
degenerate}

With $\Theta = 2\pi$, the bulk spectrum of the field theory is
identical to $\Theta = 0$, thus the disordered phase is gapped and
nondegenerate. In the 1+1d case, using explicit renormalization
group calculation, it was demonstrated that $\Theta = 2\pi$ is a
stable fixed
point~\cite{pruisken1,pruisken2,pruisken3,pruisken4,pruisken5}. In
fact, without explicit calculation, the symmetry of Eq.~\ref{3dp}
and Eq.~\ref{o3theta} determines that $\Theta = 2\pi$ must be a
fixed point which does not flow under renormalization group, while
nothing forbids other values of $\Theta$ from flowing. Thus we
will use the fixed point $\Theta = 2\pi$ to derive the edge
states.

Since the bulk is fully gapped and nondegenerate in the quantum
disordered phase when $\Theta = 2\pi$, we can safely integrate out
the bulk, and look at the boundary theory. Using the standard
bulk-boundary correspondence, we can derive the boundary theory of
Eq.~\ref{3dp}, which is a 2+1 dimensional NLSM defined in
$\mathcal{M}$ with a WZW term at level $ k = 1$: \beqn &&
\mathcal{S}_b = \int d^2x d\tau \ \frac{1}{g}
\mathrm{tr}[\partial_\mu \mathcal{P}
\partial_\mu \mathcal{P}] \cr\cr &+& \int_0^1 du \int d^2x d\tau \ \frac{i 2\pi
k}{256\pi^2} \mathrm{tr}[\mathcal{P} \partial_\mu \mathcal{P}
\partial_\nu \mathcal{P} \partial_\rho \mathcal{P}
\partial_\lambda \mathcal{P}] \epsilon_{\mu\nu\rho\lambda}.
\label{3dboundary}\eeqn Here $u \in (0, 1)$, and
$\mathcal{P}(\vec{x}, \tau, u)$ is an extension of
$\mathcal{P}(\vec{x}, \tau)$ that satisfies \beqn
\mathcal{P}(\vec{x}, \tau, 1) = \mathcal{P}(\vec{x}, \tau), \ \ \
\mathcal{P}(\vec{x}, \tau, 0) = \Omega. \eeqn The coefficient of
the WZW term in Eq.~\ref{3dboundary} must be quantized, in order
to make sure that the WZW term is a well-defined topological term
in the 2+1d field theory.

Such WZW terms can be analyzed very reliably in 0+1d and 1+1d, and
in both cases, these terms change the ground state dramatically.
In 0+1d, a WZW term leads to degenerate ground states; in 1+1d, it
drives the system to a stable gapless fixed point described by
conformal field
theory~\cite{witten1984,KnizhnikZamolodchikov1984}. In higher
dimensions, nontrivial effects of a WZW term are still expected,
but we no longer have a complete understanding. Since we are
interested in the strongly interacting disordered phase, basically
any perturbative calculation will fail, thus this is a highly
nontrivial problem. In the follows I will argue that the
disordered phase of the 2+1d boundary theory Eq.~\ref{3dboundary}
must be either gapless or degenerate.

In order to make this argument, let us first weakly break the
SU(2$N$) symmetry down to SU($N$)$\times$SU($N$)$\rtimes Z_2$.
This residual SU($N$)$\times$SU($N$) symmetry transformation can
be written as \beqn V = \left(
\begin{array}{cccc}
V_L, & \mathbf{0}_{N \times N} \\ \\
\mathbf{0}_{N \times N}, & V_R
\end{array}
\right) \eeqn while the residual $Z_2$ symmetry corresponds to
exchanging $V_L$ and $V_R$. With this symmetry reduction, the
order parameter $\mathcal{P}$ can be written as \beqn \mathcal{P}
= \left(
\begin{array}{cccc}
\cos(\theta) \mathbf{1}_{N \times N}, & i \sin(\theta) U  \\ \\
- i \sin(\theta) U^\dagger , &  - \cos(\theta) \mathbf{1}_{N
\times N}
\end{array}
\right), \eeqn where $U$ is an SU($N$) matrix. Under the
transformation $V_L$ and $V_R$, $U$ transforms $U \rightarrow
V_L^\dagger U V_R$, thus the SU($N$)$\times$SU($N$) residual
symmetry precisely corresponds to the left and right
transformation of $U$. Under the $Z_2$ symmetry transformation,
\beqn Z_2 : \ \ \theta \rightarrow \pi - \theta, \ \ \ U
\rightarrow U^\dagger. \eeqn

Under this symmetry reduction, $\theta$ and $U$ no longer have the
same energy scale. We will replace $\cos(\theta)$ and
$\sin(\theta)$ by their expectation values, and assume that the
fluctuation of $\theta$ has a higher energy scale compared with
$U$. Thus at low energy we can rewrite the boundary theory
Eq.~\ref{3dboundary} in terms of slow mode $U$. Now
Eq.~\ref{3dboundary} is reduced to a principal chiral model (PCM)
defined on manifold SU($N$) with a $\Theta^\prime$ term: \beqn
\mathcal{L}_b \rightarrow \frac{1}{g} \mathrm{tr}[\partial_\mu
U^\dagger
\partial_\mu U] + \frac{i\Theta^\prime}{24\pi^2}
\mathrm{tr}[U^\dagger
\partial_\mu U U^\dagger \partial_\nu U U^\dagger \partial_\rho U].
\label{thetap} \eeqn If the $Z_2$ symmetry of the residual
symmetry is unbroken, $i.e.$ the expectation value of
$\cos(\theta)$ is $0$, then the derived boundary SU($N$) PCM has
precisely $\Theta^\prime = \pi$. Notice that in Eq.~\ref{thetap},
$U$ is the only dynamical field. Since $U\rightarrow U^\dagger$
under the $Z_2$ transformation, $\Theta^\prime = \pi$ is a
symmetric point where the boundary Lagrangian Eq.~\ref{thetap} is
invariant under this $Z_2$ transformation. When this $Z_2$
symmetry is explicitly broken, namely at the boundary we turn on a
background field that tunes $\theta$ away from $\pi/2$, then a
straightforward calculation shows that the derived $\Theta^\prime$
is also tuned away from $\pi$, and $\Theta^\prime = 2\pi(2 +
\cos(\theta))(\sin(\theta/2))^4$. 


The phase diagram of 2+1$d$ SU($N$) PCM was studied in
Ref.~\cite{xuludwig}, where it was argued that when $g$ is large
enough, $\Theta^\prime > \pi$ and $\Theta^\prime < \pi$ are two
different disordered phases, which are both fully gapped and
nondegenerate. These two disordered phases are separated by either
a first or second order transition at $\Theta^\prime = \pi$. Thus
at $\Theta^\prime = \pi$ the system must be either gapless or two
fold degenerate. In our formalism we can see that the residual
SU($N$)$\times$SU($N$)$\rtimes Z_2$ symmetry guarantees
$\Theta^\prime = \pi$ at the boundary, $i.e.$ these symmetries
guarantee the boundary of Eq.~\ref{3dp} cannot be trivially gapped
out.

Since SU($N$)$\times$SU($N$)$\rtimes Z_2$ is a subgroup of
SU(2$N$), the original SU(2$N$) invariant model Eq.~\ref{3dp} must
also describe a $3d$ SPT. Here we did not assume any symmetry more
than SU(2$N$) or its subgroups. As we already discussed, just like
the Haldane phase, in Eq.~\ref{3dp} $\Theta = 2\pi$ is a fixed
point with a fully gapped and nondegenerate spectrum in its
disordered phase. If $\Theta$ is tuned slightly away from $2\pi$
($\Theta = 2\pi \pm \epsilon$), the bulk energy gap will not be
closed immediately, thus this perturbation can only affect the
boundary theory Eq.~\ref{3dboundary} and Eq.~\ref{thetap}.
However, $\Theta^\prime = \pi$ at the boundary theory
Eq.~\ref{thetap} is protected by the subgroup $Z_2$ of SU(2$N$),
thus as long as the spin symmetry is preserved, no perturbation
can make the edge states have a trivial spectrum. Just like the
Haldane phase, the edge state can only disappear through a phase
transition in the bulk. This argument leads to the conclusion that
in Eq.~\ref{3dp}, $\Theta = 2\pi$ is a stable fixed point of a SPT
phase. In the future it is worth to perform an RG calculation for
both $\Theta$ and $g$ in Eq.~\ref{3dp} directly, like what has
been done for the 1+1d NLSMs
\cite{pruisken1,pruisken2,pruisken3,pruisken4,pruisken5}.

Now let us create the following configuration of $\Theta^\prime$
at the 2d boundary (Fig.~\ref{fig}$b$): \beqn
\Theta^\prime(\vec{x}) &=& 2 \pi, \ \ \ \mathrm{for} \ |\vec{x}| <
R , \cr \cr \Theta^\prime(\vec{x}) &=& 0, \ \ \ \mathrm{for} \
|\vec{x}|
> R. \eeqn Or equivalently, the order parameter $\mathcal{P}$
on the two sides of the domain wall takes values $\mathcal{P} =
\pm \Omega$, for $|\vec{x}| < R$ and $|\vec{x}| > R$ respectively.
The two sides of the domain wall are conjugate to each other under
the $Z_2$ subgroup of SU(2$N$). According to our previous work
\cite{xuludwig}, in the disordered phase of Eq.~\ref{thetap}, at
the domain wall $|\vec{x}| = R$, there is a gapless
$\mathrm{SU}(N)_L \times \mathrm{SU}(N)_R $ conformal field theory
with level $k = 1$. The $\mathrm{SU}(N)_L$ and $\mathrm{SU}(N)_R$
charges move clockwise and counter-clockwise respectively along
the domain wall. Later these domain wall states will help to argue
that with the SU($N$)$\times$SU($N$)$\rtimes Z_2$ symmetry, model
Eq.~\ref{thetap} can only be realized at the boundary of a 3d
system, $i.e.$ Eq.~\ref{3dp} describes a 3d SPT phase.

{\it Step 2: argue the edge state cannot be realized as a 2d
system}

Let us take $N = 2$ as an example, and reinvestigate the domain
wall configuration in Fig.~\ref{fig}$b$. Let us couple the
SU(2)$_L$ charges to a U(1) gauge field $A_\mu \sigma^z$, which is
a spin gauge field that couples to $\sigma^z$ only. Based on the
gapless domain wall states, one can show that if a $2\pi-$flux of
$A_\mu$ is inserted at the origin $\vec{x} = 0$, the domain wall
will accumulate gauge charge 2~\cite{liuwen,levinsenthil}.

As a comparison, let us make a similar domain wall in a pure 2d
system described by the same SU(2) PCM Eq.~\ref{3dboundary} with
SU(2)$_L\times$SU(2)$_R \rtimes Z_2$ symmetry, and inside the
domain wall the SU(2)$_L$ and SU(2)$_R$ charges have opposite Hall
conductivities.
Since there is a $Z_2$ transformation connecting the two sides of
the domain wall, and the $Z_2$ symmetry exchanges SU(2)$_L$ and
SU(2)$_R$, thus the systems inside and outside the domain wall
should have {\it opposite} Hall conductivities of $A_\mu$. Since
for a 2d bosonic system without fractionalization and topological
degeneracy, the Hall conductivity can only be an even
integer~\cite{levinsenthil,luashvin,liuwen}, then inserting a
$2\pi-$flux of $A_\mu$ inside the domain wall will accumulate
gauge charges $4k$ at the domain wall, with integer $k$. This
proves that with the SU(2)$_L\times$SU(2)$_R \rtimes Z_2$
symmetry, the domain wall states at the boundary of the 3d SPT
described in Eq.~\ref{3dp} cannot be realized in a 2d system.

Without the $Z_2$ symmetry that connects the two sides of the
domain wall, the argument above would fail. For example, the 2+1
dimensional PCMs discussed in Ref.~\cite{liuwen} have no such
$Z_2$ symmetry that connects $\Theta^\prime = 2\pi$ and
$\Theta^\prime = 0$.

The same argument can be generalized to the case with $N
> 2$. Now we conclude that Eq.~\ref{3dp} describes a 3$d$ SPT
phase protected (at least) by the subgroup
SU($N$)$\times$SU($N$)$\rtimes Z_2$ of the SU($2N$) spin symmetry.
Since SU($N$)$\times$SU($N$)$\rtimes Z_2$ is a subgroup of
SU(2$N$), the original SU(2$N$) invariant model Eq.~\ref{3dp}
should also describe a $3d$ SPT phase.

Similar situation occurs in the ordinary 3d topological insulator:
the single Dirac cone at the boundary of a 3d topological
insulator cannot be realized in a 2d electron system with
time-reversal symmetry; but without the time-reversal symmetry, a
2d electron system certainly can have a single Dirac cone. We can
create a similar domain wall as Fig.~\ref{fig}$b$ at the boundary
of a 3d topological insulator, and break the time-reversal
symmetry on both sides of the domain wall oppositely, so the mass
gap $m$ of the 2d boundary Dirac fermion satisfies $m > 0$ inside
the domain wall ($|\vec{x}| < R$), and $m < 0$ outside the domain
wall ($|\vec{x}|
> R$). At the boundary of the 3d topological insulator, the two
sides of the domain wall have Hall conductivity $\pm 1/2$
respectively. Then a $2\pi-$flux inserted inside the domain wall
will accumulate charge $e$ at the domain wall. However, if such
domain wall is created in a pure 2d quantum Hall system, where the
two sides of the domain wall are connected through the
time-reversal transformation, then a $2\pi-$flux inserted inside
the domain wall will at least accumulate charge $2e$ at the domain
wall.

By coupling two copies of Eq.~\ref{3dp} together like
Eq.~\ref{2o3theta}, one can show that the theory with $\Theta =
4\pi$ can be continuously connected to $\Theta = 0$ without a bulk
transition. Thus again Eq.~\ref{3dp} describes a 3d SPT with $Z_2$
classification: there is a trivial class with $\Theta = 4\pi k$,
and a nontrivial Haldane class with $\Theta = (4k + 2)\pi$.
Using the same argument as Ref.~\cite{xuludwig},
we can derive a phase diagram for model Eq.~\ref{3dp}: The system
is topological when $\Theta \in \left( (4k + 1)\pi, (4k+3)\pi
\right)$, while the system is trivial when $\Theta \in \left(
(4k-1)\pi, (4k+1)\pi \right) $; $\Theta = (2k+1)\pi$ is the
transition, where the system is either gapless, or two fold
degenerate.

\subsection{3.2 Lattice Construction}

Now let us construct a trial spin state for Eq.~\ref{3dp} on a
lattice. Consider a SU(2$N$) antiferromagnet on a diamond lattice,
where there are two flavors on each site, and the spin of each
flavor on every site carries a self-conjugate representation of
SU(2$N$) (Fig.~\ref{fig}$c$). The spin operator can be represented
using slave fermion $f_{i,A, \alpha,a}$: \beqn S^\alpha_{i, \beta,
a} = \sum_{A = 1}^2 \frac{1}{2N} f^\dagger_{i, A, a, \alpha} f_{i,
A, a, \beta} - \frac{1}{2N} \delta_{\alpha\beta}. \eeqn Here $A =
1,2$ is a color index, $\alpha, \beta = 1, \cdots 2N$ are the
SU(2$N$) indices, $a = 1, 2$ denotes the two flavors. In order to
match the spin Hilbert space and fermion Hilbert space, again we
need to impose two constraints $\sum_{A, \alpha} f^\dagger_{i, A,
a, \alpha} f_{i, A, a, \alpha} = 2N $, and $f^\dagger_{i, A}
\rho^\mu_{AB} f_{i, B} = 0$, whose effects can be effectively
described by a dynamical compact U(1) gauge field, and a SU(2)
gauge field that couples to the color space of $f_{i, A,
a,\alpha}$~\cite{xusun2,hermele2}.

Instead of constructing the spin Hamiltonian, let us just consider
a spin state, where the slave fermion $f_{i,A, a,\alpha}$ fills a
similar mean field band structure as the Fu-Kane-Mele (FKM) model
on the diamond lattice, and the flavor index $a = 1, 2$ plays the
role of spin in the FKM model~\cite{fukane}: \beqn H_0 &=&
\sum_{<i,j>, A, a, \alpha}
 - t_{ij} f^\dagger_{i, A, a, \alpha} f_{j, A, a, \alpha} \cr\cr &+& i \lambda
\sum_{\ll i,j \gg} f^\dagger_{i, A, a, \alpha} \vec{\sigma}_{ab}
\cdot (\vec{d}^1_{ij} \times \vec{d}^2_{ij})f_{j, A, b, \alpha}.
\label{meanfield}\eeqn The spin wave function is obtained after
projecting the slave fermion mean field wave function to satisfy
the gauge constraints:
\begin{eqnarray} | G_{\mathrm{spin}} \rangle = \prod_i
\mathrm{P}(\hat{n}_i = 2N)\otimes \mathrm{P}(\hat{\rho}^\mu_i = 0)
| f_{i, A, a, \alpha} \rangle. \label{projection}
\end{eqnarray}

In Eq.~\ref{meanfield}, when $t_{ij}$ is uniform and isotropic,
there are three independent 3d Dirac fermions in the Brillouin
zone; two of the three Dirac fermions can be trivially gapped out
with an anisotropic $t_{ij}$, now we are left with one Dirac
fermion at $\vec{Q} = (0, 0, \pi)$, whose low energy Hamiltonian
reads \beqn H &=& \sum_{\alpha = 1}^{2N} \sum_A \psi^\dagger_{A,
\alpha} \vec{\Gamma} \cdot \vec{q} \psi_{A, \alpha}, \cr \cr &&
\Gamma_1 = \tau^z \sigma^x, \ \ \Gamma^2 = \tau^z \sigma^y, \ \
\Gamma_3 = \tau^y. \eeqn

Now we turn on a topological mass gap $H_1$ to the mean field
Dirac Hamiltonian: \beqn H_1 = \sum_{\alpha, A, B} m
\psi^\dagger_{A, \alpha} \Gamma_{4} \rho^z_{AB} \psi_{B, \alpha},
\ \ \Gamma_4 = \tau^x. \eeqn With this mass, the band structure of
the slave fermions with color index $A = 1$ is a topological
insulator, thus $A = 1$ fermions have massless $2d$ Dirac fermion
edge states; the slave fermion with color index $A = 2$ is a
trivial band insulator without edge states. $H_1$ also breaks the
SU(2) gauge field down to another U(1) gauge field generated by
$\rho^z$, $i.e.$ with nonzero $H_1$, the slave fermion $f$ is
coupled to two different compact U(1) gauge fields at low energy.

Notice that in the current situation we did not particularly
assume the time-reversal symmetry in our model, thus one might
expect the system to develop another mass gap $H_2 = m_2
\psi^\dagger \Gamma_5 \psi$, where $\Gamma_5 = \tau^z \sigma^z$.
Indeed, in the original FKM model, this is precisely the mass gap
that breaks the time-reversal symmetry and drives the system into
a trivial insulator. Namely, in the original FKM model, due to the
existence of $H_2$, a topological band structure and a trivial
band structure can be adiabatically connected to each other
without a bulk phase transition. However, the physical meaning of
this mass gap is a slave fermion density modulation between two
flavors and two sublattices, in the FKM model it is a spin density
wave. Since our original spin model requires a constant slave
fermion number on each flavor, then after gauge projection this
mass gap $H_2$ does not correspond to any physical order parameter
in our lattice spin model. Thus even without time-reversal
symmetry, there is no physical term one can turn on in the spin
Hamiltonian that connects the topological state to a trivial state
without a phase transition.

In the following four paragraphs we will demonstrate that this
lattice construction reproduces everything we have discussed in
the field theory analysis. To build the connection with the NLSM
Eq.~\ref{3dp}, we couple the slave fermions with the
antiferromagnetic N\'{e}el order parameter $\mathcal{P}$: \beqn
H_3 = m_3 \psi^\dagger_\alpha \Gamma_5 \psi_\beta
\mathcal{P}_{\alpha\beta}. \eeqn Both $H_1$ and $H_3$ are mass
gaps of the Dirac fermion. Now the entire Lagrangian of the Dirac
fermion can be written concisely as : \beqn \mathcal{L} &=&
\bar{\psi} \gamma_\mu \partial_\mu \psi + \bar{\psi} \mathcal{U}
\psi, \cr\cr \mathcal{U} &=& \cos(\vartheta) \Gamma_4 \rho^z +
\sin(\vartheta)\Gamma_5 \mathcal{P}. \label{3dorder}\eeqn
$\mathcal{U}$ is a unitary matrix. After integrating out the Dirac
fermions, a 3+1d WZW term for unitary matrix $U$ is
generated~\cite{abanov2000}: \beqn \mathrm{WZW}(\mathcal{U}) =
\int^1_0 du \int d^3x d\tau \frac{2\pi }{480\pi^3}
(\mathcal{U}^\dagger d\mathcal{U})^5. \label{3dwzw} \eeqn

Once we assume there is a nonzero background $H_1$ (the
coefficient $\cos(\vartheta)$ in Eq.~\ref{3dorder} is nonzero),
the $\Theta-$term in Eq.~\ref{3dp} can be precisely derived by
directly plugging $\mathcal{U} = \cos(\vartheta) \Gamma_4 \rho^z +
\sin(\vartheta)\Gamma_5 \mathcal{P}$ in this WZW term
Eq.~\ref{3dwzw}. And the derived $\Theta$ is precisely $2\pi$.

With a nonzero $H_1$ in the bulk, the boundary of the system is
described by $2N$ two-dimensional Dirac fermions with a SU(2$N$)
symmetry, and these Dirac fermions are coupled to two U(1) gauge
fields. The SU($N$) PCM at the boundary (Eq.~\ref{thetap}) can
also be directly derived using the boundary Dirac fermions, once
we break the SU(2$N$) symmetry to SU($N$)$\times$SU$(N)\rtimes
Z_2$. With the $Z_2$ symmetry, the derived PCM model has precisely
$\Theta^\prime = \pi$.

If the $Z_2$ symmetry at the boundary is further broken, a mass
term can be turned on at the boundary: $H_4 = m_4 \bar{\psi}
\Omega \psi$. Just like the ordinary 3d topological insulator,
this mass term $H_4$ drives the edge states into a quantum Hall
state with Hall conductivity $\pm 1/2$ for the two SU($N$) charges
respectively. Without topological degeneracy, a fractional Hall
conductivity can only occur at the boundary of a 3d system. At the
2d boundary, a domain wall of $m_4$ is precisely the domain wall
of $\Theta^\prime$ in Fig.~\ref{fig}$b$, and using the slave
fermions it is straightforward to show that there are nonchiral
gapless states localized at the domain wall of $m_4$.

We have demonstrated that at the mean field level, the slave
fermion construction is completely consistent with all the
predictions made by the field theory in the previous subsection.
So far we have ignored the dynamical gauge fields, which in 3+1
dimensional space-time can have a gapless photon phase. In order
to make sure the bulk is a fully gapped SPT, we need to drive the
system into the confined phase of the U(1) gauge fields.
Confinement of U(1) gauge field is driven by condensation of the
magnetic monopoles. In the free electron case, a monopole in a
topological band insulator will carry gauge charge due to the
topological $\Theta-$term in the electromagnetic response
function~\cite{qi2008}. The $\Theta-$term leads to ``oblique
confinement" after the monopoles condense~\cite{hooft,cardy}. But
in a system where charges are strongly interacting, the quantum
number of the lightest monopole, as well as the nature of its
confinement transition is not obvious, and I will leave this to
future studies. Condensate of bound state between monopole and
gauge charges in strongly interacting system is under active
studies right now~\cite{choxu,fisherfuture}, and in our current
work it is assumed that an ordinary confined phase is still
possible by condensing appropriate bound state of monopole and
gauge charges. The nontrivial edge physics of the 3d SPT will
survive the confinement, because for example the 1+1d CFT at the
domain wall in Fig.~\ref{fig}$b$ cannot be gapped out without
backscattering between left and right moving modes, $i.e.$ the
domain wall CFT is always stable unless the
SU($N$)$_L\times$SU($N$)$_R$ symmetry is spontaneously broken down
to its diagonal SU($N$) subgroup.

The trial lattice construction in this section can be tested by
directly studying the wave function of the slave fermions, after
turning on onsite gauge constraints. For example, the edge states
will exist not only at a physical boundary of the system, it will
also exist in entanglement spectrum. And to study the entanglement
spectrum, one only needs the ground state wave function, which is
what we have constructed in this section.


Just like the 1d Haldane phase, the state described above is a 3d
SPT phase {\it only} if the color singlet constraint
$f^\dagger_{i} \rho^\mu f_{i} = 0$ is strictly imposed on every
site. By contrast, if this constraint is softened, namely the
representation on every site is no longer the one in
Fig.~\ref{fig}$c$, another mass term can be added to the mean
field band structure of the slave fermion: $H_5 = m_5 \psi^\dagger
\Gamma_5 \rho^z \psi$, which will completely destroy the bulk SPT,
and gap out the edge states without degeneracy. $m_5$ corresponds
to a ``color density wave" on the lattice, which is not allowed
with the on-site color singlet constraint.

Our construction only applies to the self-conjugate representation
in Fig.~\ref{fig}$c$, which is invariant under the center $Z_{2N}$
subgroup of group SU(2$N$), while the fundamental representation
is not invariant under the center $Z_{2N}$. Thus the 3d SPT state
constructed in this work {\it cannot} be classified using the
cohomology of SU(2$N$) group, it might be classified using the
cohomology of group $\mathrm{PSU}(2N)$ = SU(2$N$)/$Z_{2N}$.


\section{4. Discussion and Summary}

In this work, we studied one class of 3d symmetry protected
topological phase, whose stability requires spin symmetry SU(2$N$)
or its subgroup SU($N$)$\times$SU($N$)$\rtimes$$Z_2$, but does not
require other discrete symmetries.
For large enough $N$, homotopy groups
$\pi_{2d}[\frac{\mathrm{U}(2N)}{\mathrm{U}(N) \times
\mathrm{U}(N)}]$ and $\pi_{2d-1}[\mathrm{SU}(N)]$ are always
$\mathbb{Z}$ for $d \geq 2$, thus our formalism and results can be
generalized to any odd spatial dimension. A $\Theta-$term can be
defined for N\'{e}el order parameter $\mathcal{P} \in
\frac{\mathrm{U}(2N)}{\mathrm{U}(N) \times \mathrm{U}(N)}$ in any
odd spatial dimension. After breaking the SU(2$N$) symmetry to
SU($N$)$\times$SU$(N)\rtimes Z_2$ symmetry, this bulk
$\Theta-$term always reduces to a boundary $\Theta^\prime-$term
with $\Theta^\prime = \pi$ in the SU($N$) PCM at the boundary.


This work is supported by the Alfred P. Sloan Foundation, the
David and Lucile Packard Foundation, Hellman Family Foundation,
and NSF Grant No. DMR-1151208. While preparing for this work, the
author became aware of another independent work on 3d
SPT~\cite{vishwanathsenthil}.

\bibliography{3dspt}

\end{document}